\documentclass{article}
\usepackage{arxiv}
\usepackage[utf8]{inputenc}
\usepackage[T1]{fontenc}
\usepackage{xcolor}
\usepackage[round]{natbib}
\definecolor{blendedblue}{rgb}{0.2, 0.2, 0.6}
\usepackage{graphicx}
\usepackage[bookmarks   = true,
            citecolor   = blendedblue,
            colorlinks  = true,
            linkcolor   = blendedblue,
            urlcolor    = blendedblue,
            citecolor   = blendedblue,
            linktocpage = false,
            hyperindex  = true]{hyperref}
\usepackage{booktabs}
\usepackage{amsfonts}
\usepackage{enumitem}
\usepackage{nicefrac}
\usepackage{amsmath, amssymb, amsthm}
\usepackage{url}
\usepackage{doi}
\usepackage{bm}
\usepackage{fontawesome}
\usepackage{mathtools}
\usepackage{lmodern}
\usepackage{siunitx}
\usepackage{booktabs}
\usepackage{etoolbox}
\usepackage{calc}
\usepackage{dsfont}
\usepackage{multirow}
\usepackage{array}
\usepackage{graphicx}
\graphicspath{{figures/}}
\usepackage{todonotes}
\usepackage{changes}

\definecolor{blendedblue}{rgb}{0.2, 0.2, 0.6}
\definecolor{forestgreen(web)}{rgb}{0.13, 0.55, 0.13}
\definecolor{darkorange}{rgb}{1.0, 0.55, 0.0}
\definecolor{emeraldblue}{HTML}{1eb5be}

\usepackage{amsfonts, amsmath, amssymb, enumerate} 

\usepackage[labelsep=period]{caption}

\renewcommand{\headeright}{}

\usepackage{parskip}
\makeatletter 
\renewcommand\@biblabel[1]{#1.} 
\makeatother

\title{Confidence Intervals for Conditional Covariances of Natural Frequencies} 

\renewcommand{\shorttitle}{Confidence Intervals for Conditional Covariances of Natural Frequencies}

\date{\today 
}

\author{
	\href{https://orcid.org/0000-0003-2256-1127}{\includegraphics[scale=0.06]{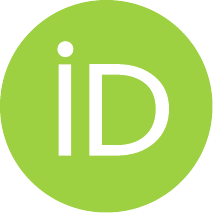}\hspace{1mm}Lizzie Neumann} \\
        Chair of Statistics and Data Science \\
 	Dept.~of Mathematics and Statistics\\
	School of Business, Economics and Social Sciences\\
    Helmut Schmidt University\\
	Hamburg, Germany\\
	\texttt{neumannl@hsu-hh.de} \\
 	\And
	\href{https://orcid.org/0000-0001-7151-8243}{\includegraphics[scale=0.06]{orcid.pdf}\hspace{1mm}Philipp Wittenberg} \\
        Chair of Statistics and Data Science \\
 	Dept.~of Mathematics and Statistics\\
	School of Business, Economics and Social Sciences\\
    Helmut Schmidt University\\
	Hamburg, Germany\\
	\texttt{pwitten@hsu-hh.de} \\
	\And
	\href{https://orcid.org/0000-0001-6777-4746}{\includegraphics[scale=0.06]{orcid.pdf}\hspace{1mm}Jan Gertheiss} \\
        Chair of Statistics and Data Science \\
 	Dept.~of Mathematics and Statistics\\
 	School of Business, Economics and Social Sciences\\
    Helmut Schmidt University\\
	Hamburg, Germany\\
	\texttt{gertheij@hsu-hh.de} \\
}

\begin{document}	

\maketitle

\begin{abstract}
    In structural health monitoring (SHM), sensor measurements are collected, and damage-sensitive features such as natural frequencies are extracted for damage detection. However, these features depend not only on damage but are also influenced by various confounding factors, including environmental conditions and operational parameters. These factors must be identified, and their effects must be removed before further analysis. However, it has been shown that confounding variables may influence the mean and the covariance of the extracted features. This is particularly significant since the covariance is an essential building block in many damage detection tools. To account for the complex relationships resulting from the confounding factors, a nonparametric kernel approach can be used to estimate a conditional covariance matrix. By doing so, the covariance matrix is allowed to change depending on the identified confounding factor, thus providing a clearer understanding of how, for example, temperature influences the extracted features. This paper presents two bootstrap-based methods for obtaining confidence intervals for the conditional covariances, providing a way to quantify the uncertainty associated with the conditional covariance estimator. A proof-of-concept Monte Carlo study compares the two bootstrap versions proposed and evaluates their effectiveness. Finally, the methods are applied to the natural frequency data of the KW51 railway bridge near Leuven, Belgium. This real-world application highlights the practical implications of the findings. It underscores the importance of accurately accounting for confounding factors to generate more reliable diagnostic values with fewer false alarms.
\end{abstract}


\section{Introduction}

Natural frequencies are known to be influenced by temperature and other confounders~\cite{Han.etal_2021,Avci.etal_2021,Wang.etal_2022}, and in structural health monitoring (SHM) various methods exist to adjust for the influence of environmental and operational parameters (EOP). Many use regression techniques, known as ``response surface modeling''~\cite{Worden.Cross_2018,Worden.etal_2016}, to model the effect of the confounders on the response, e.g., natural frequencies. Then, the predicted values are subtracted from the observed frequencies, and further analysis is based on the resulting errors, sometimes also called ``misfits''~\cite{Maes.etal_2022}. An important point to note is that the methods used for response surface modeling, ranging from simple linear regression to sophisticated machine learning, typically target only the expected value of the response. However, the analysis of real-world SHM data shows that, in particular, the temperature affects not only the mean but also higher-order statistical moments such as the (co-)variances and correlations of the system outputs, especially the natural frequencies~\cite{Neumann.etal_2025}. This is particularly harmful as the covariance is an essential building tool in many damage detection tools in SHM. Therefore, it has been proposed to replace the standard covariance matrix typically used in SHM with its conditional version, which can be estimated non-parametrically~\cite{Neumann.etal_2025,Yin.etal_2010}. This method allows the covariance matrix to vary with the confounder, e.g., the temperature, and corresponding confounder-adjusted damage detection methods (which use the conditional covariances) can be used, such as a conditional Mahalanobis distance or conditional principal component scores~\cite{Neumann.etal_2025,Neumann_2025}. This paper will present and compare two approaches to quantify the uncertainty regarding the estimated conditional covariances, namely two bootstrap methods to calculate confidence intervals for the conditional covariance. 

The remainder of the paper is structured as follows. Section~\ref{sec_CC_CI} revisits the conditional covariance estimator and presents two bootstrapping methods to obtain confidence intervals. In Section~\ref{sec_pcs}, a proof of concept Monte Carlo study will be presented, and in Section~\ref{sec_kw51}, the results obtained for the eigenfrequency data from the KW51 bridge will be discussed. Finally, Section~\ref{sec_conclusion} concludes the paper.

\section{Conditional Covariances and Confidence Intervals}\label{sec_CC_CI}

\noindent\textbf{Conditional Covariances.} Let $\mathbf{x} = (x_1,\dots, x_p)^\top\in\mathbb{R}^{p}$ be a $p$-dimensional (random) output vector, e.g., natural frequencies of $p$ different modes, $z\in\mathbb{R}$ a potential confounder, e.g., the temperature, and $\mathbf{m}(z)$ the conditional mean vector of $\mathbf{x}$ at $z$.
Furthermore, let $\bm{\Sigma}(z)$ be the \emph{conditional} covariance matrix of $\mathbf{x}$ given $z$, which can be estimated from the data using a non-parametric, Nadaraya-Watson-type 
estimator of the form \citep{Yin.etal_2010, Neumann.etal_2025}
\begin{equation}
    \label{neumann:eq_Sdef}
	\hat{\boldsymbol{\Sigma}}(z; h) = \left\{\sum_{i=1}^n K_h(z_i - z)\left[\mathbf{x}_i - \hat{\mathbf{m}}(z_i)\right]\left[\mathbf{x}_i - \hat{\mathbf{m}}(z_i)\right]^\top\right\}\left\{\sum_{i=1}^n K_h(z_i - z)\right\}^{-1},
\end{equation}
where $\mathbf{x}_i = (x_{i1}, \dots, x_{ip})^\top$, $i=1,\ldots,n$, are observations of $\mathbf{x}$, and $z_i$ is the associated confounder value. $K_h(\cdot)$ is a kernel function with bandwidth $h$ and $\hat{\mathbf{m}}(z_i)$ is an estimate of the mean of $\mathbf{x}$ at $z_i$, which can be estimated using methods such as a Nadaraya-Watson kernel estimator \citep{Yin.etal_2010, Neumann.etal_2025}, penalized regression splines \citep{Eilers.Marx_1996, Neumann.Gertheiss_2022}, local polynomial regression \citep{Cleveland.etal_2017}, or any other method for response surface modeling. The bandwidth $h$ serves as the smoothing parameter for the kernel function. A higher bandwidth results in a wider kernel and a smoother covariance function. Cross-validation can be used to find an appropriate $h$ as described in \cite{Yin.etal_2010,Neumann.etal_2025}. Based on the estimate of the conditional covariance matrix, estimates of conditional correlations can also be obtained. Open source R~\cite{R_2024} code implementing \eqref{neumann:eq_Sdef} is available on Github \cite{Neumann_2024}. 

\noindent\textbf{Confidence Intervals.} 
To obtain confidence intervals for the conditional covariances (or correlations) above, nonparametric bootstrapping can be used. Since the observations $\mathbf{x}_i$ are typically correlated over time, however, we cannot use simple random sampling from the data. To account for the (potential) correlation in the data, we use so-called \emph{block bootstrapping}, a resampling technique where blocks of data, such as entire days, are sampled instead of single instances $i$. As a modification of this approach, so-called \emph{moving block bootstrapping}, the data is not separated into non-overlapping blocks (such as days), but a window of a fixed width moves along the data. In detail, we use the following procedure for (a) block bootstrapping and (b) moving block bootstrapping~\cite{Qui.Yang_2023}, respectively.\smallskip

\begin{itemize}
    \item[1.] 
    \begin{itemize}
        \item[(a)] The dataset $\{\mathbf{x}_i,\, i = 1, \dots, n\}$ is divided into blocks $\{B_k,\, k = 1,\dots, m\}$. Each block $B_k$ contains data collected over specific periods, such as days or weeks, e.g., depending on the sampling frequency, the type of system output considered (e.g., strain measurements or natural frequencies), and the length of the measurement period. The value of $m$ represents the maximum number of available days or weeks. \smallskip
        \item[(b)] From the data $\{\mathbf{x}_i,\, i = 1, \dots, n\}$, overlapping blocks $\{B_k,\, k = 1,\dots, n-\tau\}$ are constructed, where $\tau$ is the block size, e.g., a day or a week. The $k$th block is defined by $B_k = \{\mathbf{x}_i,\, i = k,\dots, k+\tau\}$. \smallskip
    \end{itemize}
    \item[2.] These blocks are then resampled (with replacement) to create new datasets of the same size as the original dataset while preserving the temporal structure of the original data. \smallskip
    \item[3.] The conditional covariance/correlation matrix is estimated for a fine grid $z_g$, $g=1,\ldots,G$, of confounder values (e.g., temperatures). \smallskip
    \item[4.] Steps 2 and 3 are repeated $\kappa$ times to ensure that we capture a comprehensive range of variations and increase the reliability of our estimates. \smallskip
    \item[5.] An approximate, pointwise $(1-\alpha)$ confidence interval for the covariance $\sigma_{kl}(z_g)$ of $x_k$ and $x_l$ at grid point $z_g$ is obtained through $[\hat{\sigma}_{kl}(z_g)-q_{1-\alpha/2}\hat{\vartheta}_{kl}(z_g),\hat{\sigma}_{kl}(z_g)+q_{1-\alpha/2}\hat{\vartheta}_{kl}(z_g)]$, where $\hat{\sigma}_{kl}(z_g)$ is the (point) estimate of $\sigma_{kl}(z_g)$ obtained through \eqref{neumann:eq_Sdef}, and $\hat{\vartheta}_{kl}(z_g)$ is the empirical standard deviation of $\hat{\sigma}_{kl}(z_g)$ over the $\kappa$ bootstrap samples; $q_{1-\alpha/2}$ denotes the $(1-\alpha/2)$-quantile of the standard normal distribution. Confidence intervals for the conditional variances and correlation are calculated analogously. \smallskip
\end{itemize}

The confidence intervals can, for instance, be used to determine the uncertainty of the conditional covariance estimator before it is used for damage detection, e.g., using a conditional Mahalanobis distance~\cite{Neumann.etal_2025} or MEWMA control charts of conditional PCA scores~\cite{Neumann_2025}.

\section{Proof of Concept Study}\label{sec_pcs}

For proof of concept, the two bootstrapping methods (a) and (b) from Section~\ref{sec_CC_CI} are applied to artificially generated data in a Monte Carlo simulation study and compared to evaluate their effectiveness for different ground truths. 

\begin{figure}[!h]\centering
    \includegraphics[width=\linewidth]{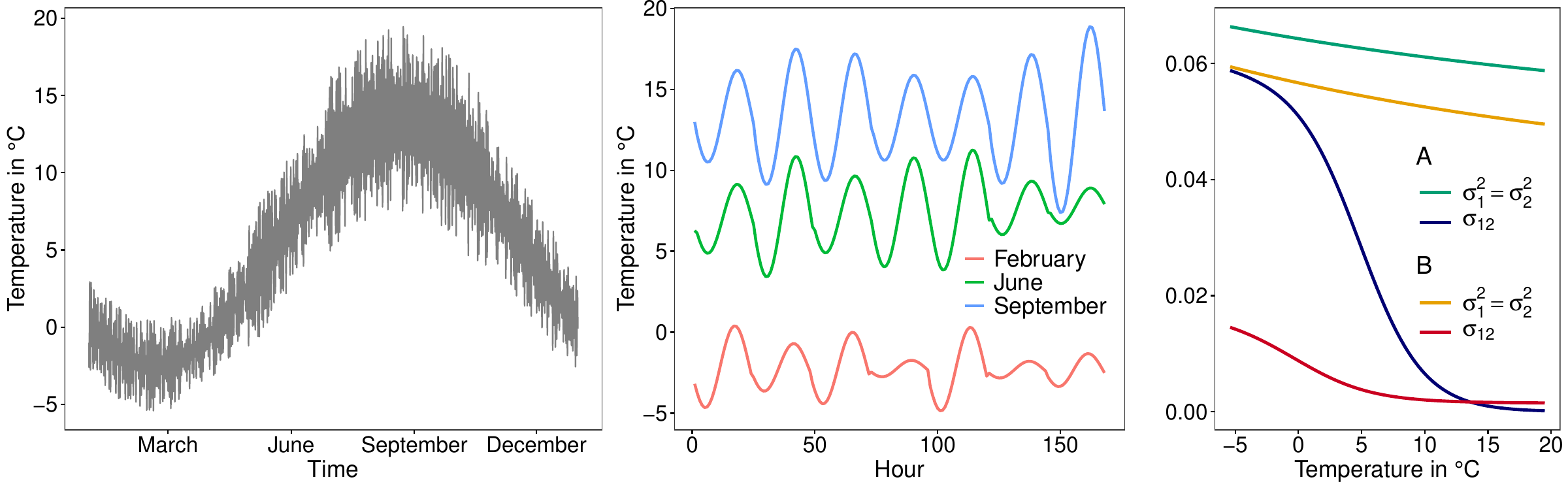}
    \caption{Illustration of the simulated temperature over the entire year (left) and the first week of three months (middle); the (true) variance functions $\sigma_1^2(z), \sigma_2^2(z)$, and covariance $\sigma_{12}(z)$ are shown in the right panel for two different scenarios, A and B.}
    \label{neumann:tempSS}
\end{figure}

We consider a latent two-dimensional normal $\mathbf{x}_t = (x_{1t}, x_{2t})^\top$ with (conditional) mean $\mathbf{m}(z) = (\mu_1(z),$ $\mu_2(z))^\top$ and (conditional) covariance matrix 
$$\Sigma(z) = 
\begin{bmatrix}
    \sigma_1^2(z) & \sigma_{12}(z)\\
    \sigma_{12}(z) & \sigma_2^2(z)
\end{bmatrix},$$ 
uncorrelated across $t \in \mathbb{Z}$. The observable system outputs are $y_{jt} = x_{jt} + \delta_{jt}$ with zero-mean, homoscedastic AR(1) error process $\delta_{jt}$, and independent across $j$. This AR(1) error process induces temporal correlation in the $y_{jt}$. The resulting covariance $\text{Cov}(y_{1t},y_{2t})$ then is $\sigma_{12}(z_t)$ from above, and $\text{Var}(y_{jt}) = \sigma_j^2(z_t) + \nu_j^2$, with $\nu_j^2 = \text{Var}(\delta_{jt})$, where $\nu_1^2 = 0.02$ and $\nu_2^2 = 0.017$ is chosen here. The temperature is modeled similarly to \cite{Wittenberg.etal_2024} to simulate typical daily and annual patterns through $z_d(\eta) = 8\text{sin}((d-141)2\pi/365) -\zeta_d \text{sin}(\pi\eta/12+0.3) + 5.5$, with day $d \in \{1,2,\ldots,365\}$, hour $\eta \in (0,24)$, and $\zeta_d \sim U(a,b)$, where $U(a,b)$ denotes the uniform distribution over the interval $(a,b)$. Four different intervals $(a,b)$ were used to ensure more variation on warmer and less on colder days. Figure~\ref{neumann:tempSS}~(left) shows the generated confounding variable ``temperature'' over the whole year, where the seasonal temperature profile is visible. Three weeks from three different months are plotted in the middle panel, visualizing the daily patterns. Two different sets of (co-)variance functions $\sigma_1^2(z), \sigma_2^2(z)$, and $\sigma_{12}(z)$ from above are considered to create two different scenarios, A and B, as shown in Figures~\ref{neumann:tempSS}~(right). 

\begin{figure}[h]
    \hspace{-3mm}
    \includegraphics[width=\linewidth]{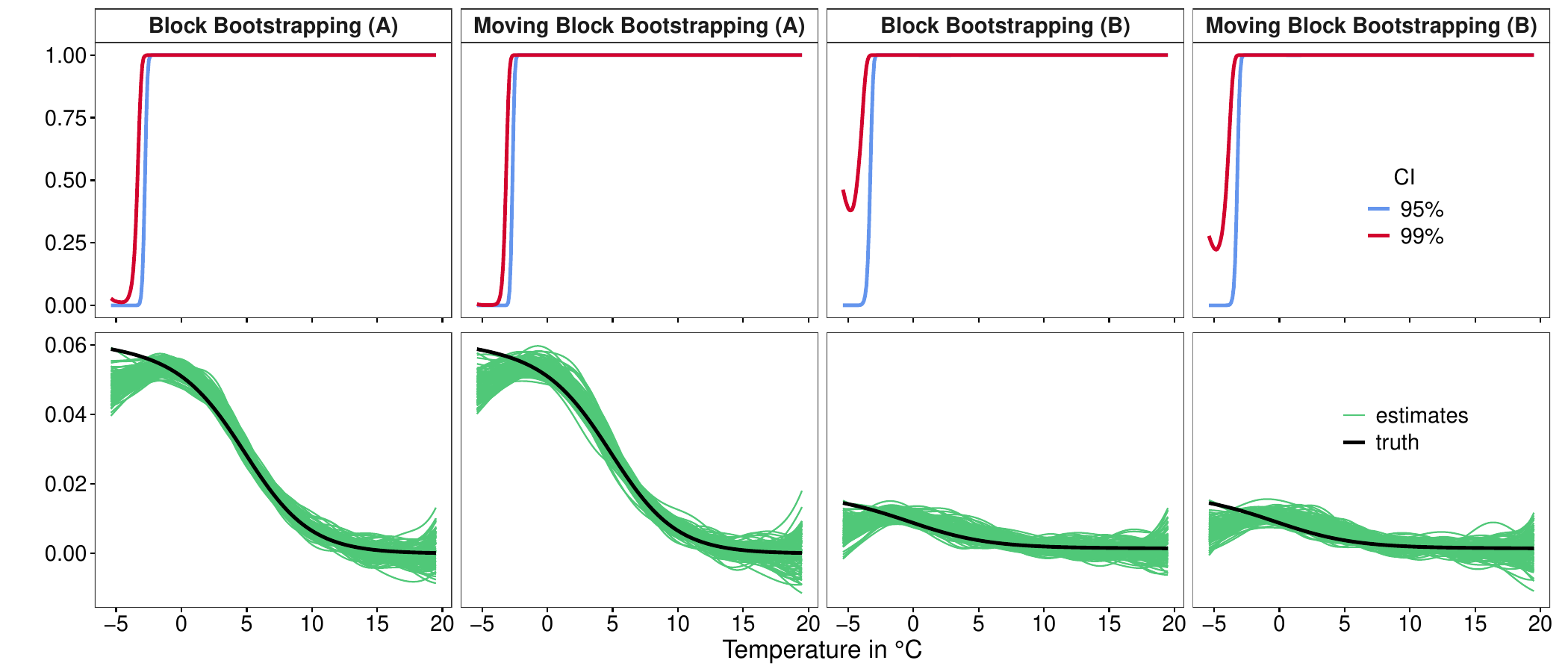}
    \caption{The coverage of the approximate, pointwise $95\%$ and $99\%$ confidences interval (top) for the covariance functions, and a sample of $100$ covariance estimates (bottom) in scenario A (left) and B (right), with the true functions in black.}
    \label{neumann:ss_coverage}
\end{figure}

The confidence intervals were estimated using $\kappa = 100$ bootstrap samples, and the estimation procedure (steps 1.--5.\ from above) was repeated on $10^4$ simulated datasets to evaluate the actual coverage of (supposedly) $95\%$ and $99\%$ (pointwise) confidence intervals of the conditional covariance functions (with the bandwidth set to $h=1.5$).
Figure~\ref{neumann:ss_coverage} shows the results for the covariance functions A (left) and B (right), respectively. The top panel shows the coverages as a function of temperature for block bootstrapping and moving block bootstrapping. Each plot in the bottom panel shows a sample of $100$ estimates of the conditional covariance (green), with the true covariance function in black for comparison.

It is seen that the coverage for temperatures above $0^\circ$C is (nearly) one, but for temperatures below $0^\circ$C, there is severe undercoverage for both the 95\% and 99\% confidence intervals in both scenarios. This can be explained by the substantial bias toward zero of the estimated covariance functions at the lower boundary around $-5^\circ$C (see Figure~\ref{neumann:ss_coverage}, bottom row). The bias is at least partly due to the limited data below $0^\circ$C. 
The overall (average) coverage of the $95\%$ and $99\%$ confidence intervals is summarized in Table~\ref{neumann:tab1}, indicating that, on average, there is only moderate undercoverage, and the proposed confidence intervals provide a helpful measure of statistical uncertainty for at least most parts of the covariance functions.
Furthermore, in both scenarios considered, block bootstrapping provided slightly better results than moving block bootstrapping. This could be seen in Figure~\ref{neumann:ss_coverage} but also from the overall coverage in Table~\ref{neumann:tab1}.
\begin{table}[h!]\small
    \centering
    \caption{Overall (average) coverage of the approximate, pointwise $95\%$ and $99\%$ confidence intervals using block bootstrapping (BB) and moving block bootstrapping (MBB).}
    \begin{tabular}{c c c c c c}
        \hline
        \multirow{ 2}{*}{Scenario} & 
        \multicolumn{2}{c}{$95\%$ confidence interval} &&
        \multicolumn{2}{c}{$99\%$ confidence interval} \\
        \cline{2-3} \cline{5-6}
          & BB & MBB && BB & MBB \\
        \cline{2-3} \cline{5-6}
       A  & $0.894$ & $0.889$ && $0.919$ & $0.910$ \\
       B  & $0.915$ & $0.912$ && $0.965$ & $0.954$ \\
       \hline
    \end{tabular}
    \label{neumann:tab1}
\end{table}

\section{The KW51 Railway Bridge}\label{sec_kw51}

\begin{figure}[!htb]
    \centering
    \includegraphics[height = 5cm]{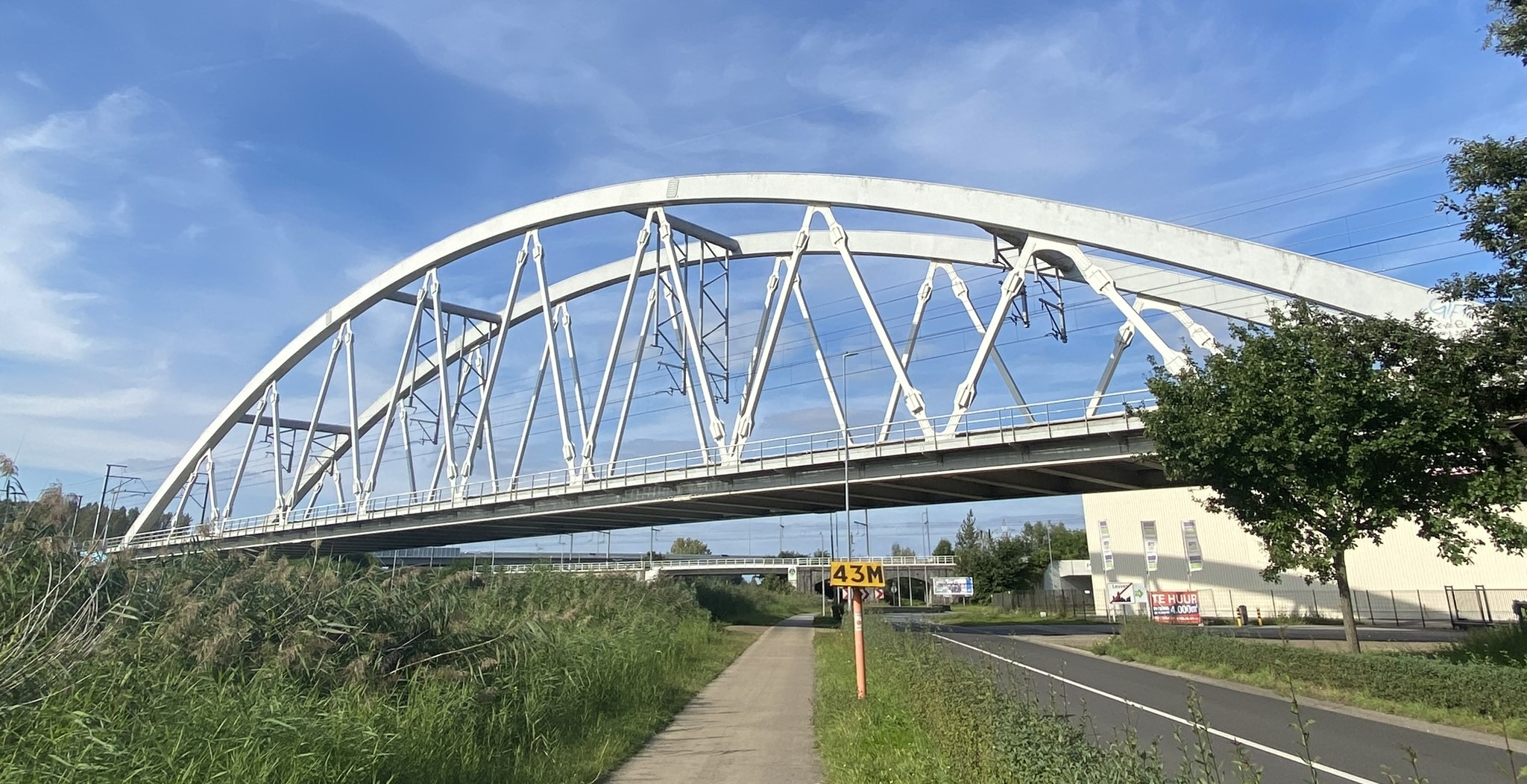} 
     \includegraphics[height = 5cm]{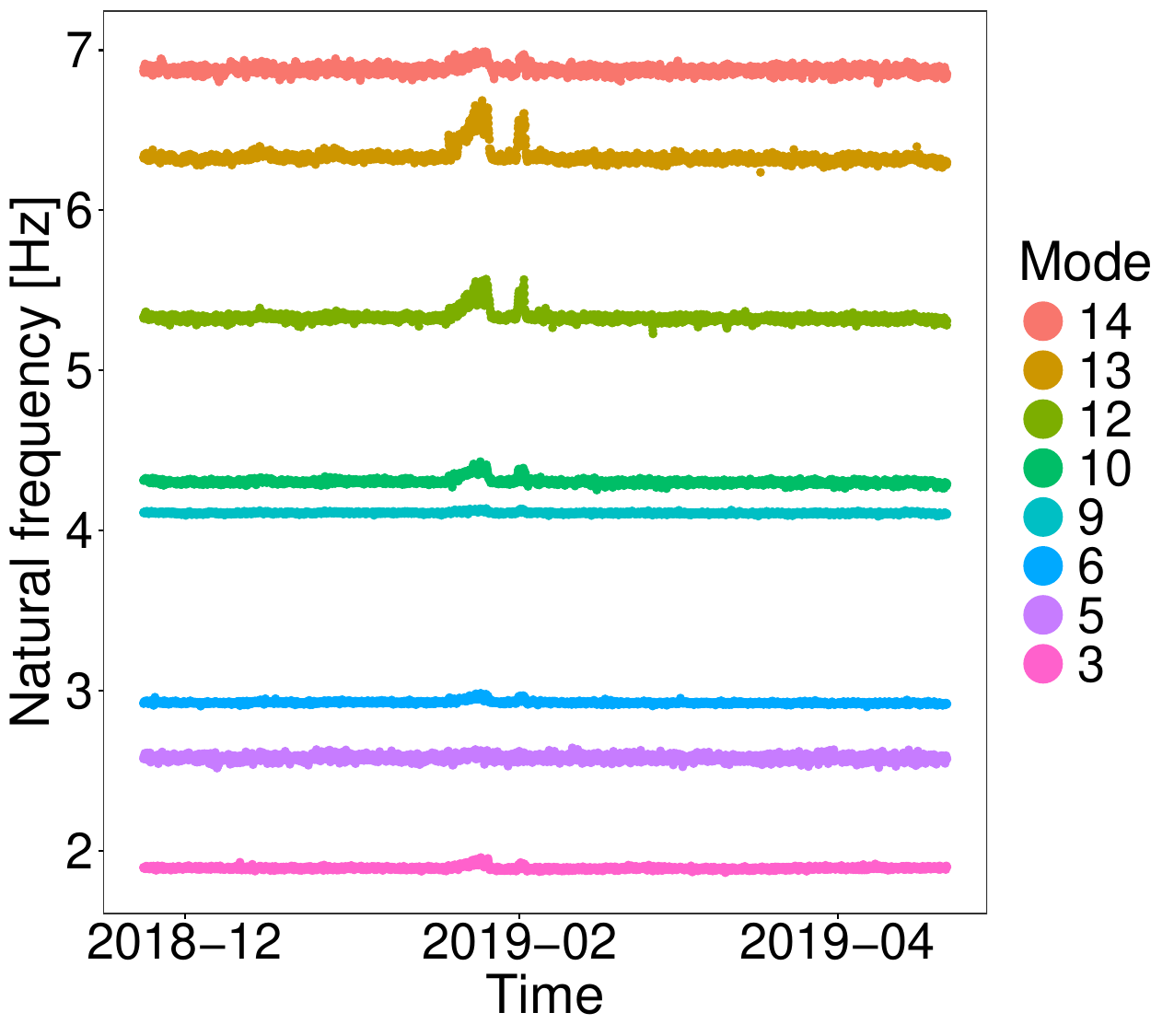} 
    \caption{KW51 bridge (in September 2024) from the south side (left) and the considered natural frequency data (right). }
    \label{neumann:fig11}
\end{figure}

The KW51 Railway Bridge~\citep{Maes.Lombaert_2021} is a steel railway bridge of the bowstring type with a length of 115 meters and a width of 12.4 meters. It consists of two curved electrified tracks and is located between Leuven and Brussels, Belgium, on the railway line L36N. The railway bridge was monitored from October 2, 2018, to January 15, 2020, with a retrofitting period from May 15 to September 27, 2019. Here, the data before the retrofitting, a period of 7.5 months, will be used. \cite{Maes.Lombaert_2021} determined the modal parameters hourly using operational modal analysis, more precisely, reference-based covariance-driven stochastic subspace identification (SSI-Cov-Ref), and tracked 14 natural frequencies over time; the steel surface temperature is also measured once per hour \citep{Maes.Lombaert_2020, Maes.Lombaert_2021, Maes.etal_2022}. 
Eight modes of vibration are used for our analysis (Modes 3, 5, 6, 9, 10, 12, 13, and 14); they are shown in Figure~\ref{neumann:fig11}~(right). The other six modes were excluded, as the modes are often not sufficiently excited and could not be tracked reliably. For the eight considered modes, missing data was supplemented using linear interpolation as done in previous studies \citep{Maes.etal_2022}.

\begin{figure}[h]
    \centering
    \includegraphics[width=\textwidth]{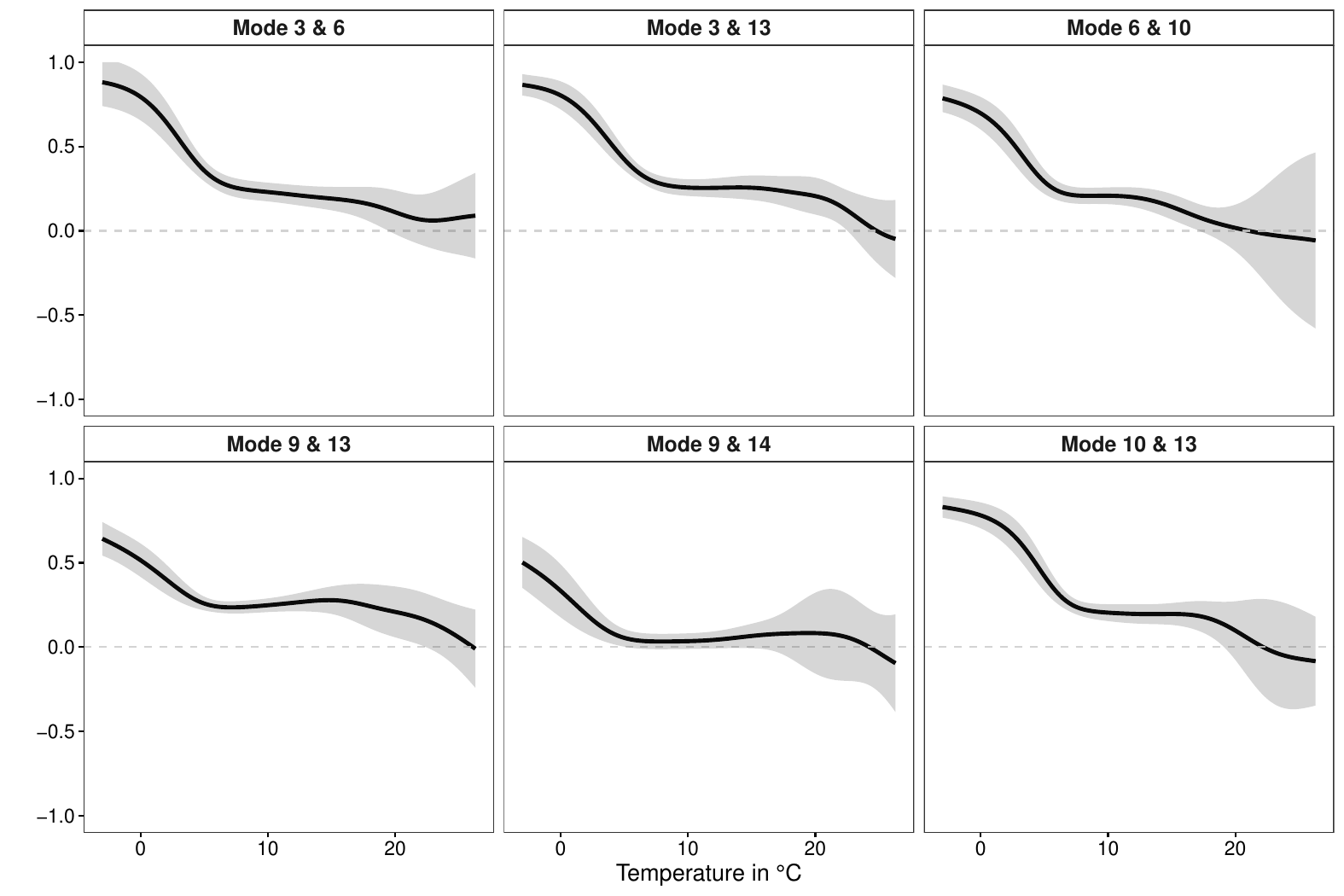}
    \caption{Pointwise 95\% confidence intervals of the conditional correlation functions for selected mode pairs of the KW51 bridge.}
    \label{neumann:ci_cest}
\end{figure}

The conditional mean was estimated using penalized regression splines \citep{Eilers.Marx_1996, Neumann.Gertheiss_2022}. They provide a smooth but nonlinear representation of the eigenfrequencies' behavior across the observed temperatures. According to \cite{Neumann.etal_2025}, the optimal bandwidth for approximately half of the mode pairs ranged from $0.6$ to $1.9$, while for the other half, it was about $0.5$. However, the results seemed robust to the specific choice of smoothing parameters. Therefore, additional smoothing was implemented by selecting a higher bandwidth. The confidence intervals were estimated utilizing block bootstrapping, and a daily sampling was chosen due to the amount of data available (approx.~$173$ days). The number of bootstrap samples was set to $\kappa = 10^5$. The resulting 95\% confidence intervals (shaded) together with the estimated correlations (solid black lines) are shown in Figure~\ref{neumann:ci_cest} as functions of the steel temperature for a selection of mode pairs. As can be seen, correlations strongly depend on the steel temperature, with the highest correlations found for temperatures below $0^\circ$C. Although, having the results of the simulation study in mind, the confidence intervals for cold temperatures may be a bit too narrow, the intervals still support the finding that there is indeed a temperature effect on correlations, e.g., in the sense that the intervals around $0^\circ$C and around $10^\circ$C do not overlap. However, estimates around $20^\circ$C and above (such as negative correlations) should be taken cautiously, as indicated by the wide intervals. The reason for the large statistical uncertainty in that region is that the data from $17^\circ$C upwards is sparse. The same is true for very low temperatures. In conclusion, having a larger amount of data substantially enhances the accuracy of estimations. The more data is available per temperature, the more reliable the results will be.

%

\section{Conclusion}\label{sec_conclusion}

This paper presents two methods utilizing block bootstrapping and moving block bootstrapping, respectively, to calculate approximate, pointwise $(1-\alpha)$ confidence intervals to assess the uncertainty of a nonparametric, kernel-based estimator for conditional covariance. The proposed methods were validated on artificially generated data in a Monte Carlo simulation study and applied to real-world SHM data, specifically the natural frequency data from the KW51 railway bridge. 
The findings regarding the KW51 bridge indicate that both the correlation and covariance significantly vary with temperature changes. Given the importance of covariances in damage detection, adjusting this metric for confounders is vital. While a confidence interval can indicate the reliability of the estimation, there are notable limitations in this method that need to be addressed. Estimating the conditional covariance requires measuring the confounder, and sufficient data must be gathered for each temperature to minimize estimation uncertainty. This issue is reflected in the KW51 natural frequencies results, where limited data for higher temperatures resulted in a wider confidence interval, indicating less dependable estimates. Conversely, when data collection is thorough, estimates prove to be reliable.
While this paper primarily focuses on natural frequencies as the relevant system output, it is also possible to estimate conditional covariances for quasi-static responses like strain or inclination. This versatility makes it a valuable tool for SHM.

\section*{Acknowledgements}
This research paper out of the project `SHM -- Digitalisierung und Überwachung von Infrastrukturbauwerken' is funded by dtec.bw -- Digitalization and Technology Research Center of the Bundeswehr, which we gratefully acknowledge. dtec.bw is funded by the European Union -- NextGenerationEU.

\end{document}